# Simulation-based Probabilistic Risk Assessment

Tarannom Parhizkar

B. John Garrick Institute for the Risk Sciences, University of California, Los Angeles, 404 Westwood Plaza, Los Angeles, CA 90095, USA; tparhizkar@ucla.edu

*SUMMARY & CONCLUSIONS*

Simulation-based probabilistic risk assessment (SPRA) is a systematic and comprehensive methodology that has been used and refined over the past few decades to evaluate the risks associated with complex systems. SPRA models are well established for cases with considerable data and system behavior information available. In this regard, multiple statistical and probabilistic tools can be used to provide a valuable assessment of dynamic probabilistic risk levels in different applications. This tutorial presents a comprehensive review of SPRA methodologies. Based on the reviewed literature, SPRA methods can be classified into three categories of dynamic probabilistic logic methods, dynamic stochastic analytical models, and hybrid discrete dynamic event and system simulation models. In this tutorial, the key strengths and weaknesses, and suggestions on ways to address real and perceived shortcomings of available SPRA methods are presented and discussed.

## 1 INTRODUCTION

While shortcomings of conventional PRA techniques have been recognized, questions have been raised regarding the capacity of the analysis methods to completely represent the interactions between the hardware, software, and human behavior [1] [1]. Moreover, the order of the events does not rely solely on the causal relationship between them, but also on their timing and the magnitude of the process variables at the time of the failure [2] [3] [4].

While shortcomings of conventional PRA techniques have been recognized, questions have been raised regarding the capacity of the analysis methods to completely represent the interactions between the hardware, software, and human behavior [1]. Moreover, the order of the events does not rely solely on the causal relationship between them, but also on their timing and the magnitude of the process variables at the time of the failure, [5] [6] [7] [8] [9].

SPRA or Dynamic based probabilistic risk assessment (DPRA) addresses the dynamic behavior of complex systems in the process of risk assessment by a systematic modeling approach, [10] [11]. As opposed to conventional PRA methods that eliminate system dynamic behavior and study system in a specific timeframe in an isolated environment, SPRA tries to understand complex and dynamic relations between system components, [12] [13].

## 2 DYNAMIC PROBABILISTIC RISK ASSESSMENT METHODS

### 2.1 Dynamic Logical Methodology (DYLAM)

A wide range of dynamic probabilistic risk assessment methods are based on probabilistic logic models. In these methods, the logic of the system behavior helps to develop the probabilistic complex system model. Based on definition, given a set E = {P(e) : e} of probabilistic examples, a background theory B, probabilistic inductive logic programming (PILP) finds, in a probabilistic logical hypothesis space $\mathfrak{H}$, [14]:

$$H^* = \arg max_{H \in \mathcal{H}} \prod_{e \in E} P(e|H, B) \quad (2)$$

Such that
$$\forall e \in E : P(e|H^*, B) \geq 0 \quad (3)$$
Where,
$$P(e|H, B) = \frac{P(H,B|e)P(e)}{P(H,B)} \quad (4)$$

The conditional probability of system behavior/status is updated based on the governing logics of the system. The dynamic logical methodology (DYLAM) is derived from logical analytical methodologies to analysis the dynamic behavior of complex systems [15]. DYLAMs generates all different possible paths resulting from an initial event. The dynamic behavior of events is reflected through time-dependent probability estimation. In [15], the general concept and features of DYLAM is presented and discussed. In this study, a powerful tool is developed that is able to integrate classic fault tree/event tree and dynamic probabilities. A simple tank case study is presented, and numerical results are discussed.

Modarres et al. [16] study is one of the examples of DYLAM method. In this study, the dependencies in multi-unit nuclear power plants are evaluated according to the available fault trees of each unit and logical dependency of units. The dependencies among multi-unit nuclear power plants have four categories of 1) causal dependence of an event (systems, structures and components (SSC) state) in one unit to another event(s) in other units 2) common (identical) SSCs shared between multiple units; 3) parametric (traditional) common cause events within one unit and across multiple-units among similar SSCs, initiating events or human errors; 4) causal dependence of an initiating event and/or SSC failures in one unit to an event external to the SSCs of other units (seismic, flood, loss of power). The proposed conceptual model identifies and explicitly models these dependencies in the modeling procedure. A simple two-unit logic is presented as an example to demonstrate the multi-unit PRA procedure proposed in this paper. Results show that all dependencies are important and should be considered in the modeling process.

In the reviewed studies, it is claim that DYLAM does not consider dependency of component failure, inspection intervals and maintenance of equipment. This method is also limited by underestimating the failure probability as a result of cutoff probability laws.

2.2 *Dynamic Flowgraph Methodology (DFM)*

Dynamic flowgraph methodology (DFM) presents the logic of the system in terms of causal relationships between physical variables of the control systems. The dynamic behavior of complex systems is represented as a series of discrete state transitions. DFM is used to identify the probability of events occurrence in a system; that can be used to identify system faults resulting from unanticipated combinations of hardware failures, software logic errors, adverse environmental conditions, and human errors. DFM has been used to assess the reliability of space rockets control systems [7], but also nuclear power plant [6] and chemical batch processes [8]. The main concept of dynamic flowgraph is presented in [17].

In 2007, ASCA [18] performed a study on probabilistic risk assessment of mission-critical software-intensive systems. In this study, software risk is modeled based on the typical PRA method integrated with DFM. The proposed methodology is applied to the mini spacecraft called Mini AERCam system that is designed and developed by the NASA Johnson Space Center.

In addition, in 2011, NASA [19] proposed a PRA procedure guideline for NASA managers and practitioners. In this guideline, it is recommended to use DFM for representation and analyzing of the dynamic interaction between system and control-software variables and parameters. It is mentioned that specific deductive analysis techniques are available to the analyst via DFM implementation software such as DYMONDATM. A case study of the Mini AERCam GN&C function utilizes DFM to consider time-dependent interaction of hardware and software components. The GN&C function utilizes input from a GPS receiver and onboard gyros. The provided inputs are then elaborated by autonomous control software algorithms to guide the thrusters to execute rotational/translational motion.

2.3 *Go-Flow*

The GO-FLOW method is a success-oriented system analysis technique. In this method, the GO-FLOW chart should be developed for the complex system. This chart is composed of operators and signal lines and represents a function of the complex system. Signal lines present the physical quantities of the system, and the intensity of a signal represents the probability of potential or actual existence of a physical quantity. In [20], a detailed explanation on this method is presented. In addition, the applications of the method on a boiling water reactor emergency core cooling system and a pressurized water reactor auxiliary feedwater system are analyzed and discussed.

2.4 *Markov Models*

Markov models composed of comprehensive representations of possible chain of events in a complex system that correspond to sequences of failures and repair. The Markov model evaluates the probability of being in a state at a given time, the amount of time that the system is expected to spend in the given state, as well as the expected number of transitions between states. This model is mostly used to calculate failures and repairs processes.

In 2015, Cicotti et al. [21] developed a DPRA model for Ambient Intelligence Healthcare Systems (AmI-HSs). The Ambient Intelligence (AmI) is a software-based system capable of supporting medical activities and procedures carried out in a high-regulated and complex healthcare environment. The model quantitively assess the risk level of different hazard scenarios in order both to support the design and development of AmI-HSs and to provide those objective evidence needed during the regulatory process. The risk model is based on a Markov decision process (MDP) that is capable of taking into account context-awareness and personalization. The application of the proposed model to a department of Nuclear Medicine is presented and results are discussed.

In 2018, Hejase et. al. [22] proposed a DPRA model for autonomous vehicle control systems. The DPRA model utilizes Markov cell-to-cell mapping technique to identify risk significant scenarios that may lead to violations of safety goals. In addition, backtracking process algorithm (BPA) is used to improve the proposed technique. The application of the BPA method on three autonomous vehicle case studies are presented and discussed.

In 2016, Yang et. al. [23] proposed a backtracking algorithm to improve Markov/cell-to-cell-mapping technique (CCMT) performance. The CCMT method is utilized to quantify the probabilistic system evolution in time, and tracing of fault propagation throughout the system. The applicability of the method is demonstrated using an example level control system and by identifying possible sequential pathways and risk significant scenarios for a given failure mode of the system. Results show that the proposed algorithm increases the fidelity of the model without increasing the required memory for execution.

### 2.5 *Dynamic Bayesian Belief Network*

Dynamic Bayesian Network (DBN) extend standard Bayesian network (BN) by defining the concept of time in the modeling. In DBN adjacent time steps relates variables of the BNs to each other over a time series or sequences. DBNs are used to model the dynamic behavior of a complex system under uncertainty. Zhao et. al. [24] used a DBN to perform fault diagnosis of a nuclear power plant. The input conditional probabilities for the DBN is relying on the expert judgment. In 2016, Jones et. al. [25] investigated the viability of severe accident management guidelines (SMART) for nuclear power plants using Bayesian networks. The developed Bayesian network is utilized to diagnosis of two types of accidents based on a comprehensive data set. Kullback-Leibler (KL) divergence method is used to gauge the relative importance of each of the plant's parameters. In this study, the accuracy and F-scores four Bayesian networks (a baseline network that ignores observation variables, a network that ignores data from the observation variable with the highest K-L score, a network that ignores data from the variable with the lowest K-L score, and a network that includes all observation variable data) are evaluated and compared.

In 2020, Groth et. al. [26] proposed a methodology which leveraged simulation, dynamic probabilistic risk assessment, and dynamic Bayesian networks to provide real-time diagnostic and monitoring for severe accidents in a nuclear power plant. In this methodology, a dynamic Bayesian network is adapted for risk management of complex engineering systems. The application of this methodology in risk-informed accident management framework, called safely managing accidental reactor transients' procedures, is discussed. In addition, a prototype model for diagnosing and monitoring system conditions associated with loss of flow and transient overpower accidents in a generic sodium fast reactor is presented and discussed.

### 2.6 *Dynamic Fault Tree (DFT) Methods*

Dynamic fault tree (DFT) utilizes dynamic gates, as well as traditional FT gates to model the dynamic behavior and interdependency of components of a complex system. The concept of dynamic fault trees is presented in multiple studies such as [27] and [28]. In 2013, Dugan et. al. [29] presented a patent for dynamic probabilistic risk assessment methodology. In this method, DFT nodes are considered as pivot nodes in PRA. The proposed mathematical method not only supports all common functions of PRA, but also include modularization, phased mission analysis, sequence dependencies, and imperfect coverage abilities.

### 2.7 *Dynamic Event Trees*

Dynamic Event Tree (DET) approach is the most commonly used one due to its direct integrability into conventional PRA. DETs could be directly integrated into existing risk models [30] [31]. In [32], the DET is coupled with fault trees to inform branching probabilities.

In 1999, Swaminathan et. al. [33] proposed a method that considered dynamic behavior of systems in event sequence diagrams (ESDs). In the presented methodology, ESDs are coupled with dynamic methodology computational algorithms which will solve the underlying probabilistic dynamics equations. This study is one of the pioneer studies in proposing a general conceptual model for dynamic event sequence diagrams. All the basic concepts and blocks of dynamic event sequence diagrams are illustrated and explained in detail.

In [34], the application of DETs in steam generator tube rapture in nuclear power plants is studied.

## 2.8 Monte Carlo Simulation (MCS)

Monte Carlo simulations are used to simulate the probability of potential outcomes in a complex system that its behavior cannot easily be predicted. In this method, instead of a value, distributions of possible inputs and any factor that has inherent uncertainty build the model, and the output is the possible results generated based on the input distributions. This method is used to model system behavior under uncertainty, and it has been used to evaluate dynamic probabilistic risk level of complex systems as well.

In 2010, Yang et. al. [35] investigated a general framework for dynamic operational risk assessment based on Monte Carlo simulation for oil/gas and chemical industries. In this framework, the relationship among scope identification and system description, hazard identification, scenario identification, and component failure mode identification are clarified. an oil/gas separator on control level is used to demonstrate the framework process. In this study, the main challenges of dynamic risk assessment methods are not discussed. In 2015, Mattenberger. et. al. [36] compared static, rapid fault-tree hybrid approach and dynamic risk assessment by providing three case studies as examples. The static risk model utilizes SAPHIRE software to build fault trees of an RCS thruster system. the model captures the probabilities of loss-of-mission (LOM) and loss-of-crew (LOC) in the system. The rapid fault tree hybrid approach utilizes the Ames Reliability Tool (ART), an Excel-based, implicit event-tree/fault-tree generator developed at NASA Ames Research Center. This model is able to capture dynamic reallocation of demands after failure by using well-known cold spare or stand-by unit redundancy calculations. The dynamic risk model utilizes Monte Carlo simulation software called GoldSim to address dynamic interactions and dependencies between all components and failure modes. The simulation not only estimates the loss-of-mission (LOM) and loss-of-crew (LOC), but also estimates loss-of vehicle (LOV) or crew-stranding at international space station (ISS). In this study, a comparison between these three models is performed and the advantages of dynamic risk assessment is highlighted.

GoldSim [37] is a software developed for dynamic probabilistic risk assessment of complex system based on Monte Carlo method. GoldSim has the ability to model the external environment, components that have multiple failure modes, complex operating rules, and complex interdependencies. This software is a general purpose and applicable to a wide variety of complex systems, e.g., mines, watersheds, waste disposal sites processing facilities, machines, space missions.

## 3 HYBRID DISCRETE DYNAMIC EVENT AND SYSTEM SIMULATION (HDDESS)

### 3.1 Accident Dynamics Simulator paired with the Information, Decision, and Action in a Crew Context Cognitive (ADS-IDAC) Model

Accident dynamics simulator paired with the information, decision, and action in a crew context cognitive model (ADS-IDAC) has gone through an evolutionally process over the past 25 years with a number of software versions [38] [39] [40] [41] [42] [43] [44] . These versions have some similarities as well as differences, both in capabilities and focus on different aspects of advanced HRA and dynamic PRA analysis. The general conceptual model of the ADS-IDAC is presented in Figure 1, [45]. The model consists of 6 main modules including user interface, scheduler, crew, indicator, system, hardware reliability modules.

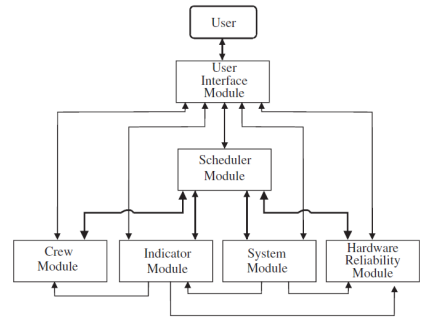

Figure 1. General conceptual model of the ADS-IDAC model

In 1996, Hsueh et. al. [46] proposed the principal modelling concepts of the accident dynamic simulator for DPRA of nuclear power plants. The simulator uses discrete dynamic event trees as the main accident scenario approach. The main modules of ADS include an operator, a plant thermal-hydraulic and safety systems response models. The capabilities of ADS are demonstrated by DPRA of the steam generator tube rupture event of a US nuclear power plant. In 1997, Smidts et. al. [40] introduced IDA model that is a cognitive model for analyzing the behavior of operators in nuclear power plants under accident conditions. The presented cognitive model is coupled with a model of control room.

In 2006, Chang et. al. [47] [48] [42] [49] [50] provided a series of five papers that discuss the information, decision, and action in crew context (IDAC) model for human reliability analysis in detail. The IDAC model predicts the behavior of the crew in nuclear power plant control rooms during an accident applicable to probabilistic risk assessment models. The crew model includes three types of operators: decision maker, action taker, and consultant. The influencing factors and flow of data in these three types of operators are presented and a general framework for IDAC is provided. The application of the model on an example is analyzed and results are discussed.

In 2007, Zhu et. al. [51] developed a methodology that integrates software contributions in the DPRA environment. The software is modeled based on multi-level objects in the simulation based DPRA environment. This research is the first systematic effort to integrate software risk contributions into DPRA models. Nejad et. al. [52] expanded scheduler module, which is one of the main modules of the ADS-IDA model. The scheduling module is responsible for generating failure scenarios in ADS-IDA. In this research a multi-level scheduling module to generated failure scenarios more efficiently is proposed. One of the main challenges of scenario generation is the problem of handling the large number of scenarios without compromising. In this study, hierarchical planning is used to generate a relatively small but complete group of risk scenarios to indicate the unsafe behaviors of the system. The scenario generation and scheduling are adjusted continuously over time by running system behavior simulation.

In 2017, Diaconeasa [53] integrated qualitative and quantitative hybrid causal logic into the ADS-IDA model. By this improvement, ADS-IDA model is able to perform analyze highly dynamic and complex accident scenarios. The investigated model includes dynamic fault trees, more advanced system and operating crew models, human failure evens (HFEs), and uncertainty propagation through the generated discrete dynamic event tree (DDET). In this research, a user-friendly graphical interface is also provided for the ADS-IDAC simulation engine. In another study, Diaconeasa [54] proposed a binary decision diagram (BDD) to connect hardware fault trees to the main event sequence diagram of the system. In this method, an algorithm for incorporating binary logic of system failures into the dynamic branching rules of the Discrete Dynamic Event Trees DDETs based on the hybrid causal logic (HCL) methodology is proposed. Using this method, hardware failure and recovery capabilities are extended to include multiple failures and recoveries during operation.

In 2019, Li et. al. [55] present new advancements in modeling of operator knowledge-based behavior in accident conditions, improving realism of the IDAC model, and simulation approach to human reliability analysis (HRA). In this study, a reasoning module is developed and implemented in ADS-IDAC to model operators' knowledge-based reasoning. The cognitive architecture of the reasoning module including knowledge representation, a memory representation, information processing flow, reasoning sequence generation, and rules for accident diagnosis is presented and explained in detail. In addition, multiple examples are presented to illustrate different features of the reasoning module.

3.2 *Monte Carlo Simulation with the Discrete Dynamic Event Tree (MCDET)*

The Monte Carlo Simulation with the Discrete Dynamic Event Tree (MCDET) is a probabilistic simulation method proposed by [56]. MCDET is an integration of Monte Carlo (MC) simulation and the discrete dynamic event tree (DDET) approach. For each set of values provided by the MC simulation, MCDET generates a new DDET. In [56], the application of this method on a German nuclear power plant is analyzed and discussed.

3.3 *Simulation Codes System for Integrated Safety Assessment (SCAIS)*

Integrated safety assessment (ISA) method utilizes the concept of a sequence as an ordered set of events to consider deterministic and probabilistic aspects of the safety problem, taking into account mutual dependencies. This method has three main steps, [57]:
1. The sequences are considered as groups of transients accounting for all uncertainties such as uncertainty in boundary conditions, parameter uncertainty, variability in occurrence time of events and initial conditions.
2. A number of sequence transients are simulated to find the sequence failure domain.
3. The exceedance frequency of failure domain is computed based on theory of stimulated dynamics (TSD) algorithms.

Simulation codes system for integrated safety assessment (SCAIS) is a computational platform of ISA analysis that consist of multiple interconnected modules as presented in Figure 2.

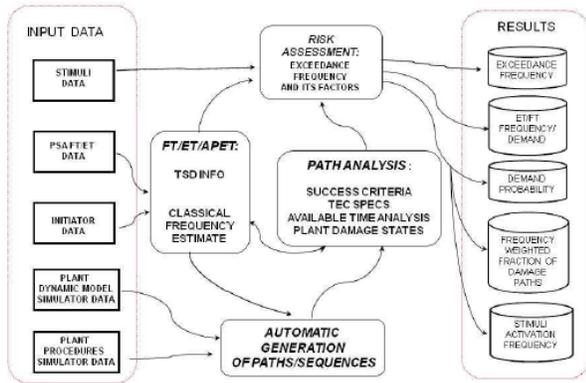

Figure 2. Simplified scheme of the SCAIS methodology [57]

Input data and results are stored in a structured query language (SQL) relational database that is accessible by all involved modules.

FT/ET/APET module consists of multiple methods and algorithms to evaluate the probability of output branches of a branching point. These probabilities are used for elimination of some of the branches on the basis of low probability termination criteria.

Path analysis module performs analysis of individual event tree sequences by simulating specific transients (paths) of the analyzed sequences. This module gets multiple simulation cases by varying values of uncertain parameters and/or time delays from automatic generation of path module and generates event tree sequences.

Risk assessment module evaluates the design barrier safety limit exceedance frequency, i.e., the failed state, by combining the TSDs over the failure domain evaluated from the path analysis module, and considering the frequency density function evaluated from the FT/ET/APET module.

The automatic generation of paths/sequences module consist of four main sub-modules:
4. The plant model that simulate accident sequences;
5. A simulator of operating procedures to implement the operator actions requested by the procedures;
6. The event scheduler that drives the dynamic generation and management of different event sequences resulting from a given initiating event;
7. A general simulation driver that combines internal and external simulation modules from which the user can configure the plant model in the form of a topology of interconnected modules.

The interconnection of these four modules are presented in Figure 3, and discussed in [57].

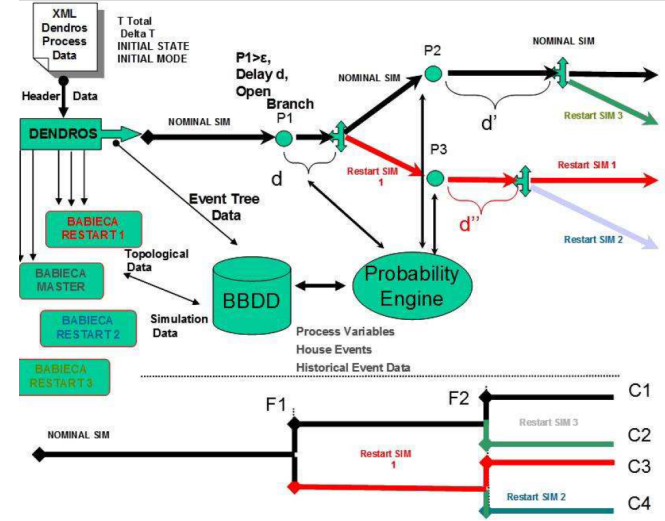

Figure 3. Path planning method in SCAIS methodology [57]

### 3.4. *Analysis of Dynamic Accident Progression Trees (ADAPT)*

Analysis of dynamic accident progression trees (ADAPT) is a method to analysis complex system dynamic risk level. Figure 4 presents ADAPT system overall architecture. As presented, ADAPT consists of three main modules including a database, a scheduler and a probability module. In the data base, all system parameters, dynamic event trees and probabilistic risk assessment databases are stored. This database is in the connection with the scheduler and probability module that perform branching and probabilistic risk assessment of the model.

The ADAPT server is in the connection with multiple human machine interfaces and exchange data. In addition, it is connected to parallel computers to perform execution in parallel, which reduces the execution time significantly.

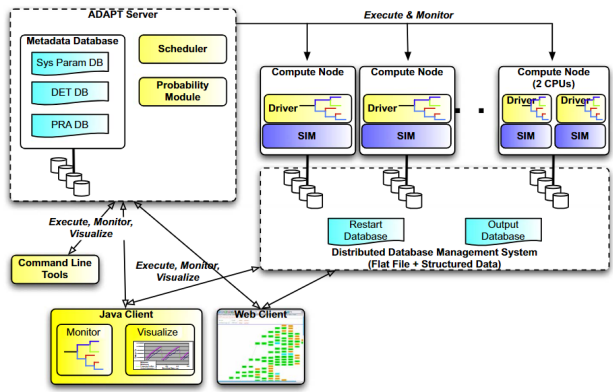

Figure 4. ADAPT system architecture [58]

In 2018, Jankovsky et. al. [59] linked the sodium-cooled fast reactor safety analysis simulator (SAS4A/SASSYS-1) to ADAPT in order to perform dynamic probabilistic risk assessment of the system. In this research, a case study is presented to illustrate the linking process. In addition, dynamic importance measures are introduced to assess the significance of reactor parameters/constituents on evaluated consequences of initiating events.

### 3.5 Reactor Analysis and Virtual Control Environment (RAVEN)

A very large effort is underway to develop a comprehensive simulation environment for risk analysis called risk analysis and virtual environment (RAVEN), developed at the Idaho National Laboratory (INL). RAVEN is a risk analysis and virtual environment that provides a framework to analyze the response of systems, employing advanced numerical techniques and stochastic algorithms. RAVEN is part of the multi-physics object-oriented simulation environment (MOOSE) [60], a parallel finite element framework for integrating fully coupled, fully implicit multi-physics solvers. The fundamental idea behind RAVEN is to employ various sampling strategies to perturb the timing and sequencing of events, initial conditions, and internal parameters of physical models to estimate the probability of occurrence of unintended events with high consequence.

In 2013, Alfonsi et. al. [61] presented RAVEN software structure and its utilization in connection with RELAP-7. Reactor excursion and leak analysis program (RELAP) is a simulation tool for safety analysis of nuclear power plants. The software is a control logic driver and post-processing tool for the thermo-hydraulic code RELAP-7. Figure 5 present RAVEN probabilistic and parametric framework. As shown, RAVEN is connected with 5 modules including a database manager, probability engine, generic model, models and post-processing and data mining module.

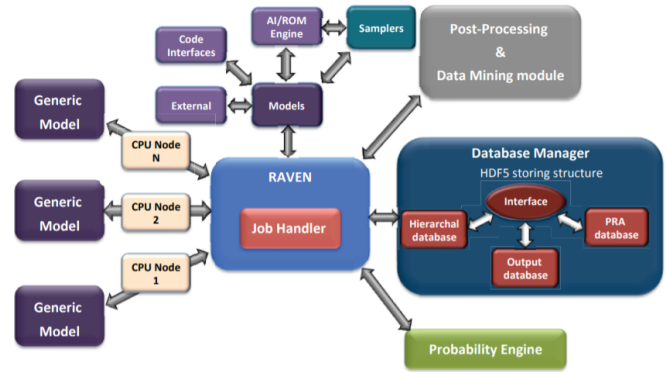

Figure 5. RAVEN probabilistic and parametric framework [61]

In [61], the mathematical model of the code is presented along with on-line controlling/monitoring and Monte-Carlo sampling. In addition, a demo of risk analysis of a station black out in a simplified pressurized water reactor (PWR) model is presented in order to illustrate the Monte-Carlo and clustering capabilities. The RAVEN consists of four modules including a RAVEN/RELAP-7 interface, a python control logic a python calculation driver, and a graphical user interface.

In 2019, Mandelli et. al. [62] assessed the dynamic probabilistic risk level of a multi-unit plant using RAVEN coupled with RELAP5-3D. The paper described and modeled the dependencies and shared resources of a multi-unit plant from both a deterministic and stochastic point of view. The coupled model is employed to assess the risk level of a site under construction with three units.

Applied to small-scale problems, this approach is a very powerful method. Nevertheless, for complex systems the set of uncertain parameters that characterize these systems grows exponentially large. Therefore, even with the fairly large computational resources that are accessible nowadays, exploring the hyperspace with a good level of confidence is generally not affordable.

*BIOGRAPHIES*

Tarannom Parhizkar, PhD, PE
B. John Garrick Institute for the Risk Sciences
University of California, Los Angeles, 404 Westwood Plaza, Los Angeles, CA 90095, USA.

e-mail: tparhizkar@ucla.edu

Dr Tarannom Parhizkar is a scientist in the B. John Garrick Institute for the Risk Sciences, at University of California, Los Angeles (UCLA). Dr. Parhizkar has worked on modeling of complex systems and performance optimization since 2010. She has an extensive background and experience both in academia and industry, addressing complex challenges encountered in risk assessment and management practices.